\documentclass[twocolumn,prl,showpacs]{revtex4} 

\usepackage[T1]{fontenc}
\usepackage{times} 

\usepackage{amsmath,amssymb,graphicx,textcomp}

\renewcommand\epsilon{\varepsilon}
\renewcommand\phi{\varphi}
\renewcommand\theta{\vartheta}
\renewcommand\vec[1]{\boldsymbol{\mathrm{#1}}}
\newcommand\unitvec[1]{\vec{\hat {#1}}}
\newcommand\diff{\mathrm{d}}
\newcommand\dotprod{\boldsymbol{\cdot}}

\newcommand\expect[1]{\left\langle\vphantom{\big(}#1\right\rangle}

\newcommand\eq[1]{Eq.~\eqref{eq:#1}}
\newcommand\fig[1]{Fig.~\ref{fig:#1}}

\newcommand\Dcm{D_\text{cm}}
\newcommand\Drot{D_\text{rot}}
\newcommand\tcoll{\tau_\text{coll}}
\newcommand\td{\tau_\text{d}}
\newcommand\tzz{\tau_\text{zz}}
\newcommand\trot{\tau_\text{rot}}

\begin{document}
\makeatletter
\def\normalsize{%
    \@setfontsize\normalsize\@xipt{12}%
    \abovedisplayskip 10\p@ \@plus2\p@ \@minus5\p@
    \belowdisplayskip \abovedisplayskip
    \abovedisplayshortskip  \abovedisplayskip
    \belowdisplayshortskip \abovedisplayskip
    \let\@listi\@listI
}%
\makeatother
\normalsize

\title{Enhanced Diffusion of a Needle in a Planar Course of Point Obstacles}
\author{Felix H{\"o}f\/ling}
\author{Erwin Frey}
\author{Thomas Franosch}
\affiliation{Arnold
Sommerfeld Center for Theoretical Physics (ASC)  and Center for
NanoScience (CeNS), Fakult{\"a}t f{\"u}r Physik,
Ludwig-Maximilians-Universit{\"a}t M{\"u}nchen, Theresienstra{\ss}e 37,
80333 M{\"u}nchen, Germany}

\begin{abstract}
The transport of an infinitely thin, hard rod in a random, dense array of point obstacles is investigated by molecular dynamics simulations. Our model mimics the sterically hindered dynamics in dense needle liquids.
Transport becomes increasingly fast at higher densities, and we observe a power-law divergence of the diffusion coefficient with exponent 0.8.
This phenomenon is connected with a new divergent time scale, reflected in a zigzag motion of the needle, a two-step decay of the velocity-autocorrelation function, and a negative plateau in the non-Gaussian parameter.
Finally, we provide a heuristic scaling argument for the new exponent.
\end{abstract}


\maketitle

Model liquids of infinitely thin, hard needles have trivial equilibrium properties, but exhibit complex dynamics.
Their equilibrium statistical mechanics is trivial, since the excluded volume is zero and all configurations are permitted and equally likely; in particular, the equation of state and the pair distribution function coincide with those of an ideal gas. But the mutual steric hindrance of moving needles gives rise to strong \emph{dynamic} correlations for dense systems.
The dynamically induced steric interaction can even generate dynamic arrest: a glass transition of needles has been observed in simulations~\cite{Renner:1995,Ketel:2005} and is rationalized within a microscopic theory~\cite{Schilling:2003+Schilling:2003a}. Due to the absence of static correlations, this glass transition is of very different origin than in structural glass formers.

The close relationship between liquids and granular gases~\cite{Goldhirsch:2003} suggests similar phenomena also for granular needles and fibers, which constitute highly anisotropic particles with convex shape.
There, interesting slow relaxation phenomena with algebraic tails were predicted~\cite{Piasecki:2006,Huthmann:1999} and are likely to be verified soon as experimental progress continues~\cite{Galanis:2006,Blum:2006}.

In early molecular dynamics simulations of a needle liquid, a surprising behavior of the center-of-mass (CM) diffusion was found by \textcite{Frenkel:1981+Frenkel:1983}, who conjectured a \emph{divergence} of the self-diffusion coefficient $\Dcm$ with increasing density $n$. In the dilute regime, the diffusion coefficient decreases as $1/n$ in accord with Enskog theory, but it starts growing beyond a crossover density. In a theoretical attempt to explain this observation, they predicted $\Dcm \sim \sqrt{n}$. An increase of $\Dcm$ by a factor~2 above its minimum value
was confirmed a few years later~\cite{Magda:1986} and also in a more recent simulation~\cite{Otto:2006}. Such a behavior contradicts the experience that transport becomes slow in dense, complex liquids.

In this Letter, we present extensive molecular dynamics simulations for a single needle exploring a planar course of impenetrable, frozen point obstacles.
This model again possesses trivial equilibrium properties, but exhibits intriguingly complex dynamics as we shall demonstrate below by a surprising enhancement of CM diffusion.
Since the needle is confined by the obstacles to a narrow tube, its motion is essentially one-dimensional leading to a strong suppression of
rotational diffusion~\cite{Needle_rotation:2008}. The same underlying physics occurs also in dense needle liquids, where the tube is formed dynamically by the surrounding needles.

\begin{figure}[b]
\includegraphics[width=.9\linewidth]{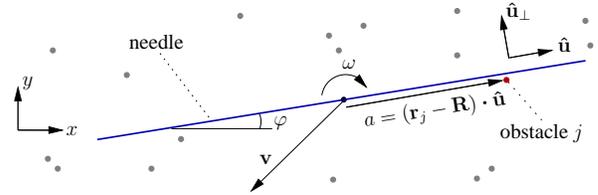}
\caption{Geometry and notation for the needle hitting an obstacle. The momentum transfer $\Delta \vec p\parallel \unitvec u_\perp$ is uniquely determined for a frictionless, elastic collision; it induces also a torque changing the angular velocity of the needle.}
\label{fig:needle_geometry}
\end{figure}

The needle is characterized by its length $L$, mass $m$, and moment of inertia $I$; we use a homogeneous mass distribution, $I=m L^2/12$. The degrees of freedom (DOFs) of the needle encompass the CM position $\vec R$ and the unit vector of orientation $\unitvec u$, parametrized by a single angle $\phi$ in our planar model; the conjugate momenta are $m \vec v$ and $I \omega$, see \fig{needle_geometry}.
We choose the collisions to be elastic and frictionless; then the total kinetic energy is conserved and its value merely sets the overall time scale of the problem. 
We distribute the obstacles randomly and independently with average number density $n$; then the only dimensionless control parameter of the model is the reduced density, $n^* := n L^2$.
A second length scale $\xi:= n^{-1/2}$ is given by the typical distance between obstacles; the crossover from the dilute to the constrained regime is expected when $\xi$ competes with $L$.

We attach an instantaneous, body-fixed, right-handed, ortho-normalized reference frame $(\unitvec u, \unitvec u_\perp)$ to the needle; it is convenient to define the two-dimensional vector product $\vec a\times \vec b:=\det(\vec a,\vec b)$, hence $\unitvec u\times\unitvec u_\perp=1$.
The laws of collision for a needle hitting a point obstacle follows from two assumptions: (i)~the collision is elastic and (ii)~the momentum transfer occurs perpendicularly to the needle, $\Delta\vec p=\Delta p\,\unitvec u_\perp$.
Simultaneously, a collision generates a torque on the needle proportional to the lever arm~$a$, given by the distance between the collision point and the axis of rotation, which yields a total change of angular momentum $a\unitvec u \times \Delta \vec p = a \Delta p$. The post-collisional velocities $\vec v_\text{f} = \vec v_\text{i} + \Delta \vec p/m$ and $\omega_\text{f} = \omega_\text{i} + a \Delta p/I$ follow from the pre-collisional ones, $\vec v_\text{i}$ and $\omega_\text{i}$, imposing conservation of kinetic energy,
\begin{equation}
\Delta p = 2mI\frac{\vec v_i \times\unitvec u  - \omega_i a}{I  + m a^2}\,.
\label{eq:collision_law}%
\end{equation}%

Our simulations of the molecular dynamics are based on an event-driven algorithm~\cite{Needle_rotation:2008,Hoefling:PhD_thesis}; the challenging part is collision detection for objects of zero width. This is solved in an efficient and robust way
by applying the interval Newton method for root finding~\cite{Hansen:IntervalAnalysis}.
It determines the zeros of the time-dependent distance, $d(t)=[\vec R(t)-\vec r_j]\times \unitvec u(t)$, between the needle and a particular obstacle at $\vec r_j$.
For each density, we have run at least 300~trajectories, each covering up to 10\textsuperscript{7}~collisions in the dense regime.

The collisions conserve the kinetic energy of the needle and redistribute it continually and equally between the two translational and one rotational DOF. The phase space trajectories are chaotic as in the case of the Lorentz model~\cite{Bunimovich:1981,Beijeren:1995}, and we have numeric evidence that the system is ergodic at all investigated densities; in particular, equipartition holds, $m\expect{\vec v^2}/2=I\expect{\omega^2}$. Furthermore, the mean collision rate follows the low-density result, $\tau_\text{coll}^{-1}\propto n^*$, also at high density;
we have measured the prefactor to $\tau_\text{coll}^{-1}=0.845\, n^*v/L$, where $v$ denotes the root mean-square velocity. In the low-density regime, the rotational and translational diffusion coefficients obey $\Drot,\Dcm\sim \tcoll$ as expected from the molecular chaos
hypothesis.


\begin{figure}
\includegraphics[width=\linewidth]{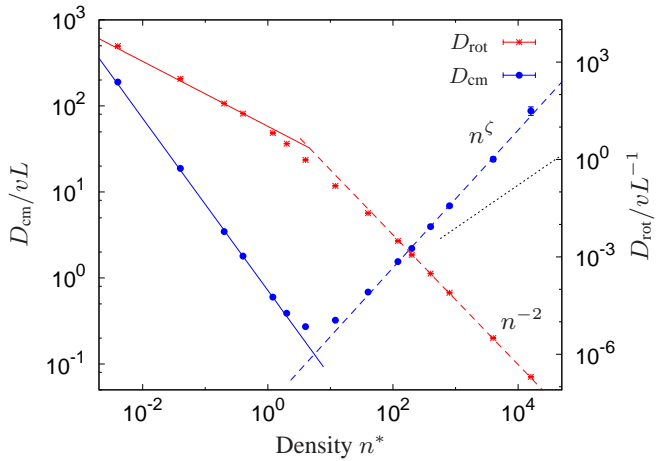}
\caption{Simulation results for the diffusion coefficients of the needle. Solid lines show the low-density behavior, $D\sim 1/n$, dashed lines are fits to the asymptotic high-density behavior, and the dotted line indicates slope 1/2 for comparison with a scaling prediction.}
\label{fig:diffusion-points}
\end{figure}

The dense array of obstacles leads to the formation of an effective tube caging the needle. Following the conventional paradigm, transport should be drastically suppressed. Indeed, the rotational diffusion coefficient obeys the predicted Doi-Edwards scaling~\cite{Doi:1978} at high densities, $\Drot\sim (n^*)^{-2}$~\cite{Needle_rotation:2008}.
While the rotational diffusion slows down with increasing obstacle density, the translational diffusion $\Dcm$ exhibits a minimum at $n^*\approx 5$ followed by a rapid increase, see \fig{diffusion-points}. Within the density range covered by our simulations, $\Dcm$ is enhanced by more than two decades. Our data suggest a power-law divergence,
\begin{equation}
\Dcm\sim (n^*)^\zeta \quad \text{for} \quad n^*\to\infty,
\end{equation}
with an exponent of $\zeta=0.80\pm0.05$.
In a recent simulation for a planar ``shish kebab'' model with soft spheres~\cite{Moreno:2004+Moreno:2004a}, also a slight increase of $\Dcm$ has been observed. The enhancement of diffusion, however, is cut off at moderate densities by the competition with the slow dynamics induced by the excluded volume~\cite{Lorentz_PRL:2006}.

\begin{figure}[b]
\includegraphics[width=3in]{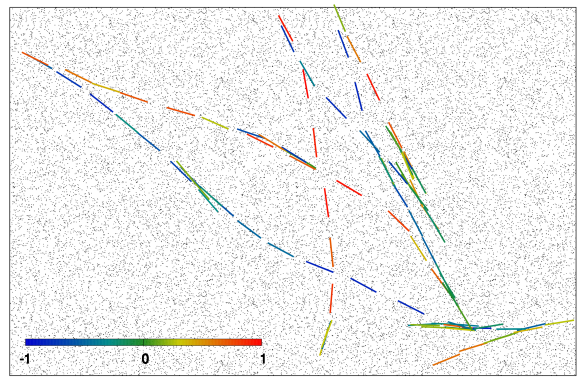} 
\caption{Snapshots of the needle at $n^*=120$. The time interval between subsequent exposures, $\Delta t=1.8~L/v$, covers about 180 collisions; colors encode the longitudinal velocity, $v_\parallel/v_\text{max}$.}
\label{fig:motion}
\end{figure}

A typical trajectory exhibits a zigzag pattern and is mainly composed of long, slightly bent segments, see \fig{motion} and the Supplementary Movie~\cite{EPAPS}; the motion of the needle carves out narrow channels.
This straight motion is occasionally interrupted and the needle rests for a short while, until it either continues its way or reverses its direction.
The latter is reflected in a change of sign of the longitudinal velocity $v_\parallel$, where we split the CM velocity into longitudinal and transverse components in the co-moving reference frame,
$\vec v=v_\parallel \unitvec u + v_\perp \unitvec u_\perp$.
Such a velocity reversal is accompanied by sharp cusps in the trajectory, and the general form of the trajectories is reminiscent of the non-holonomic dynamics of a skate on ice~
\cite{Caratheodory:1933} with its local constraints due to friction.

\begin{figure}
\includegraphics[width=\linewidth]{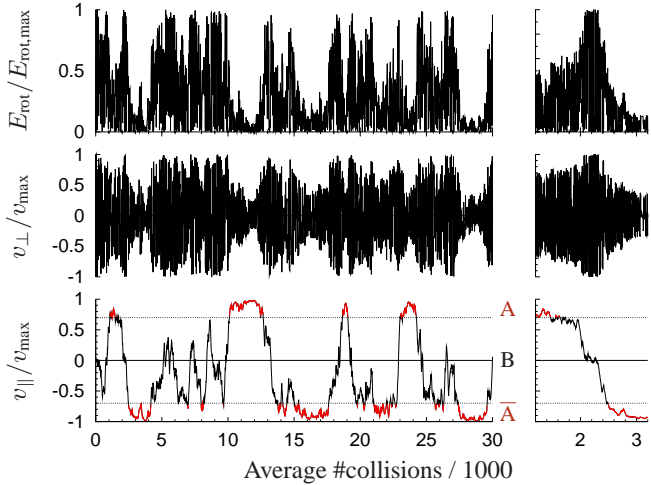}
\caption{Time series of the kinetic degrees of freedom for $n^*=120$.
Sections of a trajectory may be divided into different states: ``running'' (A), ``rattling'' (B) and ``running backwards'' ($\overline{\text{A}}$), see lower panels. The right panels enlarge a single event of velocity reversal.}
\label{fig:time_series}
\end{figure}

The time series of the kinetic DOFs in \fig{time_series} exemplifies that the transversal and rotational momenta fluctuate on the scale of individual collisions, while the longitudinal velocity fluctuates on a much slower time scale~$\tzz$. It reveals further that a reversal of~$v_\parallel$ is smeared out over many (hundred) collisions; in particular, it is not triggered by hitting a specific obstacle configuration. The dynamics may be divided into two states: ``running'' (state A) and ``rattling'' (state B).
A velocity reversal is then equivalent to switching from ``running'' (A) to ``rattling'' (B) to ``running backwards'' ($\overline{\text{A}}$). We have measured the waiting time distribution for the persistence of state~A and find an exponential distribution with rate $\tzz^{-1}$. Our analysis indicates that the events of velocity reversals realize a Poisson process  and occur independently and spontaneously.

\begin{figure}
\includegraphics[width=\linewidth]{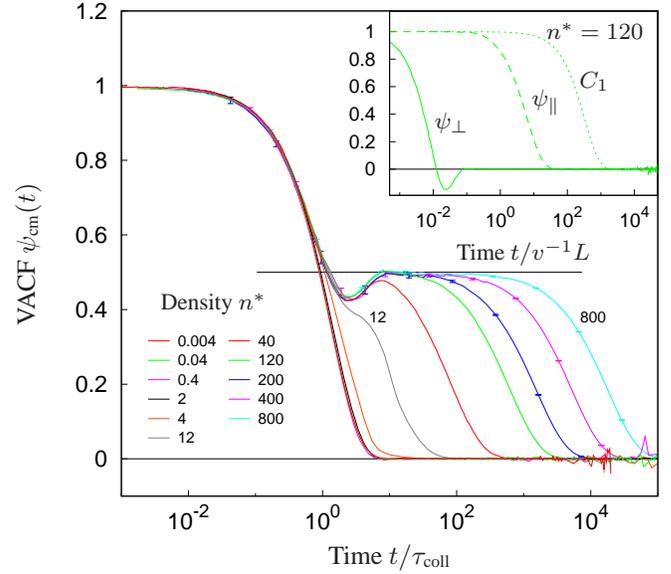}
\caption{Two-step relaxation of the CM velocity autocorrelation function $\psi_\text{cm}(t)$. Rescaling of time with the collision rate collapses the data for low densities ($n^*\leq 2$). In the dense regime, a universal intermediate plateau develops at $1/2$ for $\tcoll \ll t \ll \tzz$; it is preceded by a dip reflecting the transversal ``rattling in the cage''. Inset: correlation functions of the different DOFs entering \eq{factorization}.}
\label{fig:vacf-points}
\end{figure}

Quantitative information on the translational dynamics is obtained by considering 
the velocity autocorrelation function (VACF) $\psi_\text{cm}(t) =\expect{\vec v(t) \dotprod \vec v(0)}/ v^2$, see \fig{vacf-points}. The most prominent feature is a  two-step process for dense systems:
on the microscopic time scale, the VACF  decays towards a plateau value of $1/2$; the terminal relaxation from the plateau is exponential, providing a definition for the zigzag time $\tzz$. Thus, the initial momentum has two relaxation channels: rattling in the tube and the zigzag motion.

\enlargethispage{\baselineskip}

Let us introduce the correlation functions in the body frame $\psi_\alpha(t):=\expect{v_\alpha(t)\,v_\alpha(0)}/ \expect{v_\alpha^2}$ for $\alpha=\in\{\parallel,\perp\}$ and the orientation correlation $C_1(t)=\expect{\unitvec u(t)\dotprod \unitvec u(0)}$.
The decay rates of the three correlation functions are well separated at high densities, see inset of \fig{vacf-points}.
The strong suppression of rotational motion in particular allows to factorize the VACF,
\begin{align}
\psi_\text{cm}(t)
&= \bigl\langle \left[v_\parallel(t)\,v_\parallel(0)+v_\perp(t)\,v_\perp(0)\right] \unitvec u(t)\dotprod \unitvec u(0) \bigr\rangle / v^2 \notag \displaybreak[1] \\
&\simeq \text{\small$\dfrac{1}{2}$}\left[\psi_\parallel(t)+\psi_\perp(t)\right]C_1(t).
\label{eq:factorization}
\end{align}
We have checked numerically that this approximation is excellent for densities $n^*\gtrsim 100$.
The fastest process is the exchange of transversal momentum with the obstacles by individual collisions, $\psi_\perp(t)\approx\exp(-t/\tcoll)$;
the negative dip after 2 or 3 collisions represents the rattling of the needle in its tube. The persistent orientation yields the slowest relaxation process, $C_1(t)\simeq \exp(-t/\trot)$ where $\trot=1/\Drot$. Measuring the rate of longitudinal relaxation, $\psi_\parallel(t)\simeq \exp(-t/\tzz)$, shows an increase $\tzz\sim n^\zeta$; this provides additional evidence for the enhancement of diffusion upon exploiting a Green-Kubo relation, $\Dcm=(v^2/2)\int_0^\infty\!\psi_\text{cm}(t)\,\diff t\simeq v^2 \tzz/4$.
In summary, a ballistic needle between fixed obstacles exhibits a hierarchy of non-trivial time scales in the dense regime,
\begin{equation}
\tcoll\sim n^{-1}, \quad \td\sim n^0, \quad \tzz\sim n^\zeta, \quad \trot\sim n^2,
\end{equation}
where $\td\approx L/v$ is the disengagement time needed to leave the tube~\cite{Needle_rotation:2008}.
The scaling of all these time scales is nicely reproduced by our data; the rapid increase of $\trot$ in particular poses the computational challenge to follow individual trajectories over many decades in time.


\begin{figure}
\includegraphics[width=\linewidth]{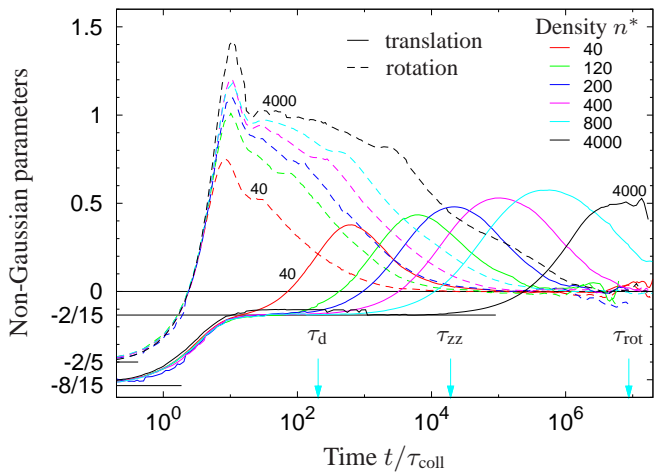}
\caption{The non-Gaussian parameters for translational and rotational DOFs, \eq{ngp}, show a rich structure at high densities including the confinement in a tube, the zigzag motion, and translation-rotation coupling. Density increases from left to right, and arrows indicate the relevant time scales for $n^*=800$.}
\label{fig:ngp-points}
\end{figure}

The interplay of translational and rotational DOFs is nicely exhibited in the non-Gaussian parameters (NGP) for CM and rotational motion, which are defined as
\begin{subequations}
\begin{align}
\alpha_2^\text{cm}(t)&=\text{\small$\dfrac{4}{3}$}\left\{\expect{|\Delta \vec R(t)|^4} \bigm/ 2 \expect{\Delta \vec R(t)^2}^2-1\right\},  \\
\alpha_2^\text{rot}(t)&=\expect{\Delta \phi(t)^4} \bigm/ 3 \expect{\Delta \phi(t)^2}^2-1.
\end{align}%
\label{eq:ngp}%
\end{subequations}
The persistent momentum renders $\alpha_2^\text{cm}(t)$ negative: after equilibration of the transversal momentum, $\alpha_2^\text{cm}(t)$ exhibits a plateau over several decades for $\tcoll\ll t\ll\tzz$. 
Noting that the component $v_\parallel$ is uniformly distributed for an ergodic sampling of the energy ellipsoid
and approximating $|\Delta \vec R(t)|\simeq v_\parallel t$, the plateau value is calculated to $-2/15$ in nice agreement with our data.
Interestingly, the plateau is followed by a positive maximum; its rising edge scales with $\tzz$, while the falling edge scales with $\trot$. Such a feature implies that the CM motion is non-trivially coupled to the rotational motion.
The orientational confinement of the needle is reflected by a peak in $\alpha_2^\text{rot}(t)$ near $t\approx\tcoll$. Then, the signal slowly degrades to a shoulder around $\td$. For longer times, $\td\ll t\ll\trot$,
the rotational NGP decays slower than $1/t$, which corresponds to an increasing Burnett coefficient---familiar from, e.g., the Lorentz model~\cite{Lorentz_LTT:2007}.

\enlargethispage{\baselineskip}

The value of the exponent $\zeta$ calls for a theoretical explanation. Let the needle initially be oriented along the $x$-axis with most kinetic energy in the longitudinal DOF, see \fig{needle_geometry}.
Since the momentum transfer per collision is perpendicular to the needle, its projection on the $x$-axis is small and fluctuates rapidly around zero, $\Delta v_x \sim \sin(\phi)$. Assuming a random walk of $v_x$, we estimate the number of collisions $N$ required for a velocity reversal to $N\sim 1/\sin(\phi)^2.$
Accepting that the collisions occur with rate $\tcoll^{-1}$, we have $\tzz\sim \tcoll N $.
At the zigzag scale $\tzz$, the angle of orientation $\phi$ diffuses, and one may approximate $\phi^2\approx 2 \Drot \tzz\ll 1$ and thus $\sin(\phi)\approx\phi$. Collecting, we obtain $\tzz\sim \tcoll/\Drot \tzz$, hence $\zeta=1/2$. This reasoning condenses the more sophisticated argument for needle liquids from Ref.~\onlinecite{Frenkel:1981+Frenkel:1983}.
Our data in \fig{diffusion-points}, however, show clearly that $\zeta=1/2$ underestimates the enhancement of diffusion.
The above argument can be modified by observing that only uncorrelated collisions contribute to $\tzz$. The mean free path to encounter an obstacle is measured by $v \tcoll$, but the needle has to traverse the distance $\xi$ to encounter a new obstacle. This suggests to use $\tzz\sim\tau_\xi N$ above, where $\tau_\xi \approx \xi/v\sim n^{-1/2}$. Thus one finds for the exponent $\zeta=3/4$, quite close to the observed $\zeta=0.8$.

In conclusion, we have observed the conjectured power-law divergence of $\Dcm$ in a 2D representation of needle liquids, thus demonstrating the relevance of our results for needle liquids and granular gases.
We have measured the exponent $\zeta$ and find its value significantly larger than anticipated.
The origin of the divergence is attributed to the zigzag motion of the needle; further, the scaling argument for the exponent $\zeta$ does not rely on the dimension and on the conservation of the single-particle energy and thus should hold also for 3D liquids.

\begin{acknowledgments}

It is a pleasure to thank R.~Schilling and  M.~Fuchs for stimulating discussions,
and we are grateful to M.~Sperl for carefully reading the manuscript.
Financial support is acknowledged from the Nanosystems Initiative Munich.

\end{acknowledgments}


\begin{thebibliography}{24}
\expandafter\ifx\csname natexlab\endcsname\relax\def\natexlab#1{#1}\fi
\expandafter\ifx\csname bibnamefont\endcsname\relax
  \def\bibnamefont#1{#1}\fi
\expandafter\ifx\csname bibfnamefont\endcsname\relax
  \def\bibfnamefont#1{#1}\fi
\expandafter\ifx\csname citenamefont\endcsname\relax
  \def\citenamefont#1{#1}\fi
\providecommand{\bibinfo}[2]{#2}

\bibitem[{\citenamefont{Renner et~al.}(1995)\citenamefont{Renner, L\"owen, and
  Barrat}}]{Renner:1995}
\bibinfo{author}{\bibfnamefont{C.}~\bibnamefont{Renner}},
  \bibinfo{author}{\bibfnamefont{H.}~\bibnamefont{L\"owen}}, \bibnamefont{and}
  \bibinfo{author}{\bibfnamefont{J.~L.} \bibnamefont{Barrat}},
  \bibinfo{journal}{Phys. Rev. E} \textbf{\bibinfo{volume}{52}},
  \bibinfo{pages}{5091} (\bibinfo{year}{1995}).

\bibitem[{\citenamefont{van Ketel et~al.}(2005)\citenamefont{van Ketel, Das,
  and Frenkel}}]{Ketel:2005}
\bibinfo{author}{\bibfnamefont{W.}~\bibnamefont{van Ketel}},
  \bibinfo{author}{\bibfnamefont{C.}~\bibnamefont{Das}}, \bibnamefont{and}
  \bibinfo{author}{\bibfnamefont{D.}~\bibnamefont{Frenkel}},
  \bibinfo{journal}{Phys. Rev. Lett.} \textbf{\bibinfo{volume}{94}},
  \bibinfo{pages}{135703} (\bibinfo{year}{2005}).

\bibitem[{\citenamefont{Schilling and
  Szamel}(2003{\natexlab{a}})}]{Schilling:2003+Schilling:2003a}
\bibinfo{author}{\bibfnamefont{R.}~\bibnamefont{Schilling}} \bibnamefont{and}
  \bibinfo{author}{\bibfnamefont{G.}~\bibnamefont{Szamel}},
  \bibinfo{journal}{Europhys. Lett.} \textbf{\bibinfo{volume}{61}},
  \bibinfo{pages}{207} (\bibinfo{year}{2003}{\natexlab{a}});
  \bibinfo{journal}{J.~Phys.: Condens. Matter} \textbf{\bibinfo{volume}{15}},
  \bibinfo{pages}{S967} (\bibinfo{year}{2003}{\natexlab{b}}).

\bibitem[{\citenamefont{Goldhirsch}(2003)}]{Goldhirsch:2003}
\bibinfo{author}{\bibfnamefont{I.}~\bibnamefont{Goldhirsch}},
  \bibinfo{journal}{Annu. Rev. Fluid Mech.} \textbf{\bibinfo{volume}{35}},
  \bibinfo{pages}{267} (\bibinfo{year}{2003}).

\bibitem[{\citenamefont{Piasecki and Viot}(2006)}]{Piasecki:2006}
\bibinfo{author}{\bibfnamefont{J.}~\bibnamefont{Piasecki}} \bibnamefont{and}
  \bibinfo{author}{\bibfnamefont{P.}~\bibnamefont{Viot}},
  \bibinfo{journal}{Europhys. Lett.} \textbf{\bibinfo{volume}{74}},
  \bibinfo{pages}{1} (\bibinfo{year}{2006}).

\bibitem[{\citenamefont{Huthmann et~al.}(1999)\citenamefont{Huthmann,
  Aspelmeier, and Zippelius}}]{Huthmann:1999}
\bibinfo{author}{\bibfnamefont{M.}~\bibnamefont{Huthmann}},
  \bibinfo{author}{\bibfnamefont{T.}~\bibnamefont{Aspelmeier}},
  \bibnamefont{and}
  \bibinfo{author}{\bibfnamefont{A.}~\bibnamefont{Zippelius}},
  \bibinfo{journal}{Phys. Rev. E} \textbf{\bibinfo{volume}{60}},
  \bibinfo{pages}{654} (\bibinfo{year}{1999}).

\bibitem[{\citenamefont{Galanis et~al.}(2006)\citenamefont{Galanis, Harries,
  Sackett, Losert, and Nossal}}]{Galanis:2006}
\bibinfo{author}{\bibfnamefont{J.}~\bibnamefont{Galanis}},
  \bibinfo{author}{\bibfnamefont{D.}~\bibnamefont{Harries}},
  \bibinfo{author}{\bibfnamefont{D.~L.} \bibnamefont{Sackett}},
  \bibinfo{author}{\bibfnamefont{W.}~\bibnamefont{Losert}}, \bibnamefont{and}
  \bibinfo{author}{\bibfnamefont{R.}~\bibnamefont{Nossal}},
  \bibinfo{journal}{Phys. Rev. Lett.} \textbf{\bibinfo{volume}{96}},
  \bibinfo{pages}{028002} (\bibinfo{year}{2006}).

\bibitem[{\citenamefont{Blum et~al.}(2006)\citenamefont{Blum, Bruns,
  Rademacher, Voss, Willenberg, and Krause}}]{Blum:2006}
\bibinfo{author}{\bibfnamefont{J.}~\bibnamefont{Blum}},
  \bibinfo{author}{\bibfnamefont{S.}~\bibnamefont{Bruns}},
  \bibinfo{author}{\bibfnamefont{D.}~\bibnamefont{Rademacher}},
  \bibinfo{author}{\bibfnamefont{A.}~\bibnamefont{Voss}},
  \bibinfo{author}{\bibfnamefont{B.}~\bibnamefont{Willenberg}},
  \bibnamefont{and} \bibinfo{author}{\bibfnamefont{M.}~\bibnamefont{Krause}},
  \bibinfo{journal}{Phys. Rev. Lett.} \textbf{\bibinfo{volume}{97}},
  \bibinfo{pages}{230601} (\bibinfo{year}{2006}).

\bibitem[{\citenamefont{Frenkel and Maguire}(1981)}]{Frenkel:1981+Frenkel:1983}
\bibinfo{author}{\bibfnamefont{D.}~\bibnamefont{Frenkel}} \bibnamefont{and}
  \bibinfo{author}{\bibfnamefont{J.~F.} \bibnamefont{Maguire}},
  \bibinfo{journal}{Phys. Rev. Lett.} \textbf{\bibinfo{volume}{47}},
  \bibinfo{pages}{1025} (\bibinfo{year}{1981});
  \bibinfo{journal}{Mol. Phys.} \textbf{\bibinfo{volume}{49}},
  \bibinfo{pages}{503} (\bibinfo{year}{1983}).

\bibitem[{\citenamefont{Magda et~al.}(1986)\citenamefont{Magda, Davis, and
  Tirrell}}]{Magda:1986}
\bibinfo{author}{\bibfnamefont{J.~J.} \bibnamefont{Magda}},
  \bibinfo{author}{\bibfnamefont{H.~T.} \bibnamefont{Davis}}, \bibnamefont{and}
  \bibinfo{author}{\bibfnamefont{M.}~\bibnamefont{Tirrell}},
  \bibinfo{journal}{J.~Chem. Phys.} \textbf{\bibinfo{volume}{85}},
  \bibinfo{pages}{6674} (\bibinfo{year}{1986}).

\bibitem[{\citenamefont{Otto et~al.}(2006)\citenamefont{Otto, Aspelmeier, and
  Zippelius}}]{Otto:2006}
\bibinfo{author}{\bibfnamefont{M.}~\bibnamefont{Otto}},
  \bibinfo{author}{\bibfnamefont{T.}~\bibnamefont{Aspelmeier}},
  \bibnamefont{and}
  \bibinfo{author}{\bibfnamefont{A.}~\bibnamefont{Zippelius}},
  \bibinfo{journal}{J. Chem. Phys.} \textbf{\bibinfo{volume}{124}},
  \bibinfo{pages}{154907} (\bibinfo{year}{2006}).

\bibitem[{\citenamefont{H{\"o}f\/ling et~al.}(2008)\citenamefont{H{\"o}f\/ling,
  Munk, Frey, and Franosch}}]{Needle_rotation:2008}
\bibinfo{author}{\bibfnamefont{F.}~\bibnamefont{H{\"o}f\/ling}},
  \bibinfo{author}{\bibfnamefont{T.}~\bibnamefont{Munk}},
  \bibinfo{author}{\bibfnamefont{E.}~\bibnamefont{Frey}}, \bibnamefont{and}
  \bibinfo{author}{\bibfnamefont{T.}~\bibnamefont{Franosch}},
  \bibinfo{journal}{Phys. Rev. E} \textbf{\bibinfo{volume}{77}},
  \bibinfo{pages}{060904(R)} (\bibinfo{year}{2008}).

\bibitem[{\citenamefont{H{\"o}f\/ling}(2006)}]{Hoefling:PhD_thesis}
\bibinfo{author}{\bibfnamefont{F.}~\bibnamefont{H{\"o}f\/ling}}, Ph.D. thesis,
  \bibinfo{school}{Ludwig-Maximilians-Universit{\"a}t M{\"u}nchen}
  (\bibinfo{year}{2006}), \bibinfo{note}{{ISBN:} 978-3-86582-426-4}.

\bibitem[{\citenamefont{Hansen and Walster}(2004)}]{Hansen:IntervalAnalysis}
\bibinfo{author}{\bibfnamefont{E.}~\bibnamefont{Hansen}} \bibnamefont{and}
  \bibinfo{author}{\bibfnamefont{G.~W.} \bibnamefont{Walster}},
  \emph{\bibinfo{title}{Global {O}ptimization {U}sing {I}nterval {A}nalysis}}
  (\bibinfo{publisher}{Marcel Dekker}, \bibinfo{address}{New York},
  \bibinfo{year}{2004}), \bibinfo{edition}{2nd} ed.

\bibitem[{\citenamefont{Bunimovich and Sinai}(1981)}]{Bunimovich:1981}
\bibinfo{author}{\bibfnamefont{L.}~\bibnamefont{Bunimovich}} \bibnamefont{and}
  \bibinfo{author}{\bibfnamefont{Y.}~\bibnamefont{Sinai}},
  \bibinfo{journal}{Commun. Math. Phys.} \textbf{\bibinfo{volume}{78}},
  \bibinfo{pages}{479} (\bibinfo{year}{1981}).

\bibitem[{\citenamefont{van Beijeren and Dorfman}(1995)}]{Beijeren:1995}
\bibinfo{author}{\bibfnamefont{H.}~\bibnamefont{van Beijeren}}
  \bibnamefont{and} \bibinfo{author}{\bibfnamefont{J.~R.}
  \bibnamefont{Dorfman}}, \bibinfo{journal}{Phys. Rev. Lett.}
  \textbf{\bibinfo{volume}{74}}, \bibinfo{pages}{4412} (\bibinfo{year}{1995}).

\bibitem[{\citenamefont{Doi and Edwards}(1978)}]{Doi:1978}
\bibinfo{author}{\bibfnamefont{M.}~\bibnamefont{Doi}} \bibnamefont{and}
  \bibinfo{author}{\bibfnamefont{S.~F.} \bibnamefont{Edwards}},
  \bibinfo{journal}{J.~Chem. Soc., Faraday Trans.~2}
  \textbf{\bibinfo{volume}{74}}, \bibinfo{pages}{560} (\bibinfo{year}{1978}).

\bibitem[{\citenamefont{Moreno and Kob}(2004{\natexlab{a}})}]{Moreno:2004+Moreno:2004a}
\bibinfo{author}{\bibfnamefont{A.~J.} \bibnamefont{Moreno}} \bibnamefont{and}
  \bibinfo{author}{\bibfnamefont{W.}~\bibnamefont{Kob}},
  \bibinfo{journal}{Europhys. Lett.} \textbf{\bibinfo{volume}{67}},
  \bibinfo{pages}{820} (\bibinfo{year}{2004}{\natexlab{a}});
  \bibinfo{journal}{J.~Chem. Phys.} \textbf{\bibinfo{volume}{121}}, \bibinfo{pages}{380}
  (\bibinfo{year}{2004}{\natexlab{b}}).

\bibitem[{\citenamefont{H{\"o}f\/ling et~al.}(2006)\citenamefont{H{\"o}f\/ling,
  Franosch, and Frey}}]{Lorentz_PRL:2006}
\bibinfo{author}{\bibfnamefont{F.}~\bibnamefont{H{\"o}f\/ling}},
  \bibinfo{author}{\bibfnamefont{T.}~\bibnamefont{Franosch}}, \bibnamefont{and}
  \bibinfo{author}{\bibfnamefont{E.}~\bibnamefont{Frey}},
  \bibinfo{journal}{Phys. Rev. Lett.} \textbf{\bibinfo{volume}{96}},
  \bibinfo{pages}{165901} (\bibinfo{year}{2006}).

\bibitem[{EPA()}]{EPAPS}
 \bibinfo{note}{{S}ee EPAPS Document No. [to be inserted]. For more
  information on EPAPS, see http://www.aip.org/pubservs/epaps.html.}

\bibitem[{\citenamefont{Carath{\'e}odory}(1933)}]{Caratheodory:1933}
\bibinfo{author}{\bibfnamefont{C.}~\bibnamefont{Carath{\'e}odory}},
  \bibinfo{journal}{Z.~Angew. Math. Mech.} \textbf{\bibinfo{volume}{13}},
  \bibinfo{pages}{71} (\bibinfo{year}{1933}).

\bibitem[{\citenamefont{H{\"o}f\/ling et~al.}(2007)\citenamefont{H{\"o}f\/ling
  and Franosch}}]{Lorentz_LTT:2007}
\bibinfo{author}{\bibfnamefont{F.}~\bibnamefont{H{\"o}f\/ling}} \bibnamefont{and}
  \bibinfo{author}{\bibfnamefont{T.}~\bibnamefont{Franosch}},
  \bibinfo{journal}{Phys. Rev. Lett.} \textbf{\bibinfo{volume}{98}},
  \bibinfo{pages}{140601} (\bibinfo{year}{2007}).

\end{thebibliography}

\end{document}